\documentclass[11pt,onecolumn]{article}
\usepackage{a4wide}
\usepackage{mathrsfs}
\usepackage{latexsym,bm}
\usepackage{graphicx}
\usepackage{subfigure}
\usepackage{indentfirst}
\usepackage{slashed}
\usepackage{amsmath}
\usepackage{amssymb}
\usepackage{color}
\usepackage{hyperref}
\usepackage{epsfig}
\usepackage[titletoc]{appendix}
\usepackage{multirow}%
\usepackage{rotating}
\usepackage{cite}
\usepackage{ulem} 
\usepackage[top=2.9cm,bottom=2.5cm,left=2.8cm,right=3cm]{geometry}
\usepackage{tabularx}
\newcommand{\mL}{\mathcal{L}}

\newcommand{\cO}{\mathcal{O}}
\newcommand{\bra}{\langle}
\newcommand{\ket}{\rangle}
\newcommand{\nn}{\nonumber}

\newcommand{\uao}{U_A(1)}

\begin{document}
\thispagestyle{empty}
\title{ \Large \bf Updated study of the $\eta$-$\eta'$ mixing and the thermal properties of light pseudoscalar mesons at low temperatures }
\author{\small Xiao-Wei Gu, Chun-Gui Duan and Zhi-Hui Guo\thanks{zhguo@hebtu.edu.cn}  \\[0.5em]
{ \small\it  Department of Physics and Hebei Advanced Thin Films Laboratory, } \\
{\small\it Hebei Normal University,  Shijiazhuang 050024, China}
}
\date{}

\maketitle

\begin{abstract}

Within the framework of the $U(3)$ chiral perturbation theory, we revisit the masses, decay constants and the mixing parameters of the light pseudoscalar mesons $\pi, K, \eta$ and $\eta'$. The low energy constants up to next-to-next-to-leading order are determined by including the light-quark mass dependences of the various quantities from different lattice QCD simulations and relevant phenomenological inputs. Then we study the finite-temperature behaviors of the masses of the light pseudoscalar mesons. The thermal behaviors of the $\eta$-$\eta'$ mixing angles in singlet-octet and quark-flavor bases are also explored. While the masses of the $\pi, K$, and $\eta$ are increased when increasing the temperatures, the mass of the $\eta'$ turns out to be slightly decreased in the low-temperature region.

\end{abstract}

\section{Introduction}

Spontaneous breaking of chiral symmetry and the $\uao$ anomaly are two characteristic features of QCD in vacuum. The former gives the octet of the pseudo-Nambu-Goldstone bosons (pNGBs), i.e. $\pi, K, \eta$, while the latter provides an explanation to the large mass of the $\eta'$. The study of these features in hot medium is important to advance the knowledge of the QCD phase diagrams, which play crucial roles in understanding the intricate phenomena of heavy ion collisions, conducted for example at RHIC and LHC (ALICE).

Chiral symmetry restoration above the critical temperature ($T_c$) is one of the compelling signals expected at these large experimental facilities. Although at sufficiently high temperatures it is well established that the anomalous breaking of the $\uao$ symmetry will be restored~\cite{Gross:1980br,Pisarski:1983ms,Kapusta:1995ww,Schafer:1996wv}, the situation around the $T_c$ region is yet unclear and needs to be further clarified. The influence of the chiral symmetry restoration on the recovery of broken $\uao$ symmetry is also of great interest and has been the focus of many recent  works~\cite{Horvatic:2007qs,Kwon:2012vb,Fejos:2016hbp,Escribano:2000ju,Jiang:2012wm,Jiang:2015xqz,Meggiolaro:2014eua,Benic:2011fv,Contrera:2009hk,Lee:2013es,Nicola:2016jlj,GomezNicola:2017bhm,Nicola:2018vug,Bazavov:2012qja,Dick:2015twa,Sharma:2016cmz,Tomiya:2016jwr}.

The thermal behaviors of the topological and chiral susceptibilities serve as useful theoretical objects to discriminate different patterns of the $\uao$ and chiral symmetry restoration and have been extensively investigated by many lattice simulations~\cite{Bazavov:2012qja,Dick:2015twa,Sharma:2016cmz,Tomiya:2016jwr} and effective theory studies~\cite{Escribano:2000ju,Jiang:2012wm,Jiang:2015xqz,Meggiolaro:2014eua,Benic:2011fv,Contrera:2009hk,Lee:2013es,Nicola:2016jlj,GomezNicola:2017bhm,Nicola:2018vug}. The masses of the light flavor pseudoscalar mesons $\pi$, $K$, $\eta$ and $\eta'$ will be definitely affected by the restoration of the $\uao$ and chiral symmetries. Therefore to study the temperature dependences of the masses of $\pi$, $K$, $\eta$ and $\eta'$ constitutes an important approach to reveal the interplay  of the $\uao$ and chiral symmetry restoration. We mention that the conclusions on the thermal behaviors of the light pseudoscalar masses are still controversial, specially for $\eta$ and $\eta'$. E.g., in order to explain the increased abundance of the $\eta'$ in hot medium, it is concluded that the mass of the $\eta'$ needs to be reduced around 200~MeV in Ref.~\cite{Csorgo:2009pa}, which is also supported by the phenomenological study in Ref.~\cite{Benic:2011fv}. In contrast, some results from the lattice simulations and effective field theories in Refs.~\cite{Jiang:2015xqz,Dick:2015twa,Fejos:2016hbp} show that there is no obvious drop of the $\eta'$ mass even in the relatively high temperature region above $T_c$.
The puzzling problem has intrigued many studies in both lattice simulations and phenomenological discussions~\cite{Horvatic:2007qs,Kwon:2012vb,Fejos:2016hbp,Escribano:2000ju,Jiang:2012wm,Jiang:2015xqz,Meggiolaro:2014eua,Benic:2011fv,Contrera:2009hk,Lee:2013es,Nicola:2016jlj,GomezNicola:2017bhm,Nicola:2018vug,Bazavov:2012qja,Dick:2015twa,Sharma:2016cmz,Tomiya:2016jwr}.

In this work we proceed the study within the $U(3)$ chiral perturbation theory ($\chi$PT), with $\pi$, $K$, $\eta$ and $\eta'$ as its active degrees of freedom. $U(3)$ $\chi$PT provides a unified theoretical framework to simultaneously incorporate the QCD $\uao$ anomaly, spontaneous and explicit chiral symmetry breaking. The finite-temperature effects enter $\chi$PT through the chiral loops and the low energy constants (LECs) of the local operators are independent of the temperatures~\cite{Gasser:1986vb,Gerber:1988tt}. Therefore the unknown LECs can be determined by using the experimental data and lattice simulations at zero temperature. The thermal behaviors of the light pseudoscalar mesons will be then the pure predictions of $\chi$PT. We mention that one should not expect to obtain precise descriptions of the physical quantities up to sufficiently high temperature region within $\chi$PT. The reason is that at high enough temperature above $T_c$ quarks and gluons are the relevant degrees of freedom, which are not explicitly included in $\chi$PT. Nevertheless, there is strong evidence that hadronic states still play quite important roles around and below $T_c$~\cite{Shuryak:2004tx}. We focus on the thermal properties of the $\pi$, $K$, $\eta$ and $\eta'$ at low temperatures in this work.

In Ref.~\cite{Guo:2015xva}, one of the authors performed a complete next-to-next-to-leading-order (NNLO) calculation of the masses and decay constants of the light pseudoscalar mesons within $U(3)$ $\chi$PT. In the present work we first update the previous study by taking into account several independent lattice simulation results, especially the sophisticated lattice simulations in Ref.~\cite{Ottnad:2017bjt}, where the corrections of the unphysical strange quark mass and finite lattice spacing  for the $\eta$-$\eta'$ mixing parameters are performed. New lattice results, including the pion-mass dependences of the mixing angles and decay constants in the quark-flavor basis, are considered, comparing with the study in Ref.~\cite{Guo:2015xva}. After the determination of the $U(3)$ $\chi$PT LECs, we extend the discussions at zero temperature to the thermal medium with finite temperatures.  The thermal properties of the light pseudoscalar mesons will be studied.

The article is organized as follows.  The relevant chiral Lagrangians and the NNLO calculations of the masses, decay constants and mixing parameters of the pNGBs $\pi$, $K$, $\eta$, and $\eta'$ are briefly discussed in Sec.~\ref{sec.lagrangian}, where we also determine the $\chi$PT LECs by taking into account the various lattice simulation results and phenomenological inputs. The temperature behaviors of the  pNGBs will be discussed in Sec.~\ref{sec.finitetemp}. A short summary and conclusions shall be given in Sec.~\ref{sec.conclusion}.

\section{Chiral Lagrangians and the determination of the LECs}\label{sec.lagrangian}

The massive singlet state $\eta_0$, mostly responsible for the physical $\eta'$ meson,  can be systematically included in $\chi$PT within the framework of large $N_C$ QCD. Based on its argument, the quark loop induced $\uao$ anomaly effect is $1/N_C$ suppressed, indicating that the mass squared $M_0^2$ of the singlet $\eta_0$ also behaves as $1/N_C$ in the large $N_C$ limit~\cite{ua1nc}. As a result, the joint expansions of momentum squared ($p^2$), light quark masses ($m_q$) and $1/N_C$ provide the consistent power counting scheme for $U(3)$ $\chi$PT~\cite{HerreraSiklody:1996pm,Kaiser:2000gs}. For later convenience, the joint expansion scheme shall be denoted as $\delta$ expansion, being $\cO(\delta) \sim \cO(p^2) \sim \cO(m_q) \sim \cO(1/N_C)$. The relevant chiral Lagrangians and calculations of the masses, decay constants and the $\eta$-$\eta'$ mixing up to NNLO in the $\delta$ expansion have been discussed in detail in Ref.~\cite{Guo:2015xva}. In order to setup the notations, we simply recapitulate the main results here.

The $U(3)$ $\chi$PT Lagrangian at leading order (LO) in the $\delta$ expansion, i.e. $\cO(\delta^0)$, reads
\begin{eqnarray}\label{eq.laglo}
\mL^{(\delta^0)}= \frac{ F^2}{4}\bra u_\mu u^\mu \ket+
\frac{F^2}{4}\bra \chi_+ \ket
+ \frac{F^2}{12}M_0^2 X^2 \,,
\end{eqnarray}
with the basic chiral building tensors
\begin{eqnarray}\label{defbb}
&& U =  u^2 = e^{i\frac{ \sqrt2\Phi}{ F}}\,, \qquad \chi = 2 B (s + i p) \,,\qquad \chi_\pm  = u^\dagger  \chi u^\dagger  \pm  u \chi^\dagger  u \,,
\qquad    X= \log{(\det U)}\,,  \nn\\
&& u_\mu = i u^\dagger  D_\mu U u^\dagger \,, \qquad  D_\mu U \, =\, \partial_\mu U - i (v_\mu + a_\mu) U\, + i U  (v_\mu - a_\mu) \,,
\end{eqnarray}
and the $U(3)$ matrix of the pNGBs
\begin{equation}\label{phi1}
\Phi \,=\, \left( \begin{array}{ccc}
\frac{1}{\sqrt{2}} \pi^0+\frac{1}{\sqrt{6}}\eta_8+\frac{1}{\sqrt{3}} \eta_0 & \pi^+ & K^+ \\ \pi^- &
\frac{-1}{\sqrt{2}} \pi^0+\frac{1}{\sqrt{6}}\eta_8+\frac{1}{\sqrt{3}} \eta_0   & K^0 \\  K^- & \overline{K}^0 &
\frac{-2}{\sqrt{6}}\eta_8+\frac{1}{\sqrt{3}} \eta_0
\end{array} \right)\,.
\end{equation}
$F$ stands for the pion decay constant at LO in the $\delta$ counting. The last operator in Eq.~\eqref{eq.laglo} parameterizes the QCD $\uao$ anomaly effect and gives the singlet $\eta_0$ the LO mass $M_0$.

The relevant $U(3)$ $\chi$PT Lagrangians to the present study at NLO and NNLO  are~\cite{Kaiser:2000gs,HerreraSiklody:1996pm,Bijnens:1999sh}
\begin{eqnarray}\label{eq.lagnlo}
\mL^{(\delta)} = L_5 \bra  u^\mu u_\mu \chi_+ \ket
+\frac{ L_8}{2} \bra  \chi_+\chi_+ + \chi_-\chi_- \ket
+\frac{F^2\, \Lambda_1}{12}   D^\mu X D_\mu X  -\frac{F^2\, \Lambda_2}{12} X \bra \chi_- \ket\,,
\end{eqnarray}
and
\begin{eqnarray}\label{eq.lagnnlo}
\mL^{(\delta^2)} =&& L_4 \bra  u^\mu u_\mu \ket \bra \chi_+ \ket
+ L_6 \bra  \chi_+ \ket  \bra  \chi_+ \ket
+ L_7 \bra  \chi_- \ket  \bra  \chi_- \ket
+  L_{18} \bra u_\mu \ket \bra u^\mu \chi_+ \ket
+  L_{25} X \bra \chi_+ \chi_- \ket
\nn \\ &&
+ \frac{F^2\, v_2^{(2)}}{4} X^2 \bra \chi_+ \ket
+ C_{12} \bra   h_{\mu\nu}h^{\mu\nu}  \chi_+   \ket
+ C_{14} \bra u_\mu  u^\mu \chi_+ \chi_+  \ket
+ C_{17} \bra u_\mu  \chi_+ u^\mu \chi_+  \ket
\nn \\ &&
+ C_{19} \bra \chi_+ \chi_+ \chi_+  \ket
+ C_{31} \bra  \chi_-  \chi_- \chi_+ \ket\,.
\end{eqnarray}
We refer to Ref.~\cite{Guo:2015xva} for detailed discussions of the previous Lagrangians.

\begin{figure}[htbp]
   \centering
   \includegraphics[width=0.4\textwidth,angle=-0]{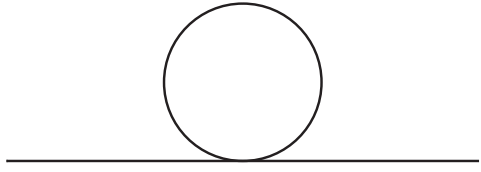}
  \caption{ Tadpole Feynman diagram for self energies of the light pNGBs. }
   \label{fig.feynmandiagram}
\end{figure}

The chiral loops start to contribute at NNLO in the $\delta$ expansion. When restricting to the masses, decay constants and the  $\eta$-$\eta'$ mixing, the only relevant chiral loops, shown in Fig.~\ref{fig.feynmandiagram}, are the tadpole functions $A_0(m_i^2)$ with different masses $m_i$ running in the loops. Following the conventional $\overline{MS}-1$ dimensional regularization scheme in $\chi$PT~\cite{Gasser:1984gg}, the expression for $A_0(m^2)$ reads
\begin{eqnarray} \label{eq.a0}
 A_0(m^2) = - m^2\ln{\frac{m^2}{\mu^2}} \,.
\end{eqnarray}
It is noted that the coefficient of $1/16\pi^2$ has been factored out in order to match the convention of  Ref.~\cite{Guo:2015xva}.

The calculations of the $\eta$-$\eta'$ mixing, the masses and decay constants of the $\pi, K$ mesons resemble the discussions in Ref.~\cite{Guo:2015xva}. As mentioned in Ref.~\cite{Guo:2011pa}, it is convenient to use the LO diagonalized fields $\overline{\eta}$ and $\overline{\eta}'$, instead of the octet $\eta_8$ and singlet $\eta_0$, when calculating the chiral loops involving $\eta$ and $\eta'$ mesons. This is due to the fact that the mixing strength between $\eta_8$ and $\eta_0$ starts from the leading order in the $\delta$ counting. For $\overline{\eta}$ and $\overline{\eta}'$, which are already diagonalized at LO, their mixing only get contributions from higher order effects that at least belong to NLO. In order to obtain the physical $\eta$ and $\eta'$ states, it is easy to work in the $\overline{\eta}$ and $\overline{\eta}'$ bases. Their general bilinear terms up to NNLO can be written as
\begin{eqnarray}\label{eq.lagmixingpara}
\mL=&&   \frac{\delta_1}{2} \,\partial_\mu\partial_\nu \overline{\eta} \partial^\mu\partial^\nu \overline{\eta}
+\frac{\delta_2}{2} \,\partial_\mu\partial_\nu \overline{\eta}' \partial^\mu\partial^\nu \overline{\eta}'
+\delta_3 \,\partial_\mu\partial_\nu \overline{\eta} \partial^\mu\partial^\nu \overline{\eta}'
+\frac{1 + \delta_{\overline{\eta}} }{2}\partial_\mu \overline{\eta} \partial^\mu\overline{\eta}
\nn \\ & &
+\frac{1+ \delta_{\overline{\eta}'} }{2}\partial_\mu \overline{\eta}' \partial^\mu \overline{\eta}'
+ \delta_k\, \partial_\mu \overline{\eta} \partial^\mu \overline{\eta}'
-\frac{m_{\overline{\eta}}^2 + \delta_{m_{\overline{\eta}}^2} }{2} \overline{\eta}\, \overline{\eta}
- \frac{m_{\overline{\eta}'}^2 + \delta_{m_{\overline{\eta}'}^2 } }{2} \overline{\eta}' \overline{\eta}'
- \delta_{m^2} \,\overline{\eta}\, \overline{\eta}' \,,
\end{eqnarray}
where the coefficients $\delta_i's$ only receive contributions from NLO and NNLO. Through the field redefinition, one can eliminate the higher derivative terms in Eq.~\eqref{eq.lagmixingpara}. Next the physical $\eta$ and $\eta'$ states can be obtained by first diagonalizing and normalizing the kinematical terms and then diagonalizing the mass terms. The relations between the physical states and the singlet-octet basis can be written in the popular two-mixing-angle form
\begin{eqnarray} \label{twoanglesmixing08}
 \left(
 \begin{array}{c}
 \eta   \\
 \eta' \\
 \end{array}
 \right) = \frac{1}{F}\left(
                                        \begin{array}{cc}
                                          F_8\, \cos{\theta_8}  & -F_0 \,\sin{\theta_0}  \\
                                           F_8\,\sin{\theta_8} & F_0 \,\cos{\theta_0} \\
                                        \end{array}
    \right)
      \left(
       \begin{array}{c}
       \eta_8   \\
       \eta_0  \\
       \end{array}
        \right)\,.
\end{eqnarray}
Alternatively, one can also relate the physical $\eta,\eta'$ states with the quark-flavor basis $\eta_q,\eta_s$, 
\begin{eqnarray} \label{twoanglesmixingqs}
 \left(
 \begin{array}{c}
 \eta   \\
 \eta' \\
 \end{array}
 \right) = \frac{1}{F}\left(
                                        \begin{array}{cc}
                                          F_q\, \cos{\theta_q}  & -F_s \,\sin{\theta_s}  \\
                                           F_q\,\sin{\theta_q} & F_s \,\cos{\theta_s} \\
                                        \end{array}
    \right)
      \left(
       \begin{array}{c}
       \eta_q   \\
       \eta_s  \\
       \end{array}
        \right)\,,
\end{eqnarray}
where the quark-flavor basis is related to the singlet-octet one through
\begin{equation}\label{etaqstoeta08}
\left(\begin{array}{c} \eta_q \\ \eta_s \end{array}\right) \quad=\quad
\left(\begin{array}{cc} \sqrt{\frac{1}{3}} &  \sqrt{\frac{2}{3}} \\  -\sqrt{\frac{2}{3}} & \sqrt{\frac{1}{3}} \end{array}\right) \quad
\left(\begin{array}{c} \eta_8 \\ \eta_0\end{array}\right) \,.
\end{equation}
Since the quark-flavor and singlet-octet bases relate each other through an orthogonal transformation, the descriptions in the two bases give equivalent results to the $\eta$-$\eta'$ mixing.  However it was noticed that the two different mixing angles $\theta_q$ and $\theta_s$ in the quark-flavor basis are quite similar and hence assumed to be equal in the so-called FKS formalism~\cite{Feldmann:1998vh}. Another way to further understand the FKS assumption has been given in Ref.~\cite{Guo:2015xva}, where it is pointed out that the FKS assumption is in accord with  neglecting of the kinematic mixing terms of the $\eta_q$ and $\eta_s$ states. The FKS assumption seems supported by many phenomenological studies and lattice simulations~\cite{Beisert:2001qb,Gerard:2004gx,Degrande:2009ps,Mathieu:2010ss,Ottnad:2017bjt}.
The mixing parameters $F_0,F_8,\theta_0,\theta_8$ in Eq.~\eqref{twoanglesmixing08} or $F_q,F_s,\theta_q,\theta_s$ in Eq.~\eqref{twoanglesmixingqs} and other quantities, such as the masses of the pNGBs and the $\pi, K$ decay constants, are given by the $U(3)$ $\chi$PT LECs and the tadpole functions. The final results are rather lengthy and have been given in detail in Ref.~\cite{Guo:2015xva}.

In this work we shall first make an update determination of the LECs by taking into account the recent lattice simulation results from ETMC~\cite{Ottnad:2017bjt}, where the $\eta$ and $\eta'$ mixing parameters and their masses have been determined by considering the corrections of the strange quark mass $m_s$ and also the finite lattice space $a$. To be more specific, we include the corrected data for the masses of $\eta$ and $\eta'$, the mixing angles in the quark-flavor basis and the ratios of $F_q/F_\pi$ and $F_s/F_K$ from Ref.~\cite{Ottnad:2017bjt}. In addition, we also consider in the fits the lattice simulations of  $m_K$, $F_\pi$ and $F_K$ from RBC/UKQCD~\cite{Aoki:2010dy,Arthur:2012yc}, and the ratios of $F_K/F_\pi$ from BMW~\cite{Durr:2010hr}. For the mixing angles in the quark-flavor basis, $\theta_q=\theta_s$ is assumed in the lattice simulation~\cite{Ottnad:2017bjt}. Within reasonable ranges of the LECs, our general mixing formalism indeed gives quite similar results for $\theta_q$ and $\theta_s$. In order to incorporate the lattice data in our study, we simply use the averages of $\theta_q$ and $\theta_s$ to fit the mixing angles  in the quark-flavor basis from Ref.~\cite{Ottnad:2017bjt}.
Other lattice results of the $\eta, \eta'$ masses from Ref.~\cite{ukqcd12prd} (UKQCD), Ref.~\cite{rbcukqcd10prl} (RBC/UKQCD), Ref.~\cite{hsc11prd} (HSC) are also used in the fits.

In the LO Lagrangian~\eqref{eq.laglo}, there are two unknown parameters $F$ and $M_0$. At LO, the decay constants of the pion and kaon are degenerate and equal to $F$. As noticed in Ref.~\cite{Guo:2015xva}, the LO fit already leads to reasonable  descriptions of the masses of $\eta$ and $\eta'$ with just one free parameter $M_0$. We closely follow the former reference to perform the LO fit, by using the masses of the  $\eta$ and $\eta'$ from the revised lattice simulations in Ref.~\cite{Ottnad:2017bjt}. We shall take the same inputs for the physical masses as those in Ref.~\cite{Guo:2015xva}. The LO mass for the singlet $\eta_0$ is 
\begin{eqnarray}\label{eq.lom0}
 M_0= 820.0 \pm 7.6~{\rm MeV}\,,
\end{eqnarray}
which is close to the value $M_0= 835.7 \pm 7.5$ in Ref.~\cite{Guo:2015xva}.

In the NLO Lagrangian~\eqref{eq.lagnlo}, four unknown LECs appear: $L_5, L_8, \Lambda_1$ and $\Lambda_2$, which will be fitted to several independent sets of lattice simulation data. In the meantime, one can also determine $F$ at NLO, by taking into account the lattice simulations of the $\pi, K$ decay constants. It will be shown later that the NLO fits reasonably reproduce the lattice data related to the $\eta$-$\eta'$ mixing, including the masses, the mixing angles and the ratios of $F_q/F_\pi$ and $F_s/F_K$ from Ref.~\cite{Ottnad:2017bjt}. However, the fit quality of the $\pi, K$ decay constants is still poor at NLO, as shown later. In order to simultaneously analyze the lattice simulations of the light pseudoscalar mesons, we need to introduce the NNLO contributions to improve the fits.

For the NNLO Lagrangian~\eqref{eq.lagnnlo}, eleven additional free LECs enter. It is not possible to obtain stable fits with so many free parameters by only including the $\eta$-$\eta'$ mixing and the decay constants of pion and kaon. For the five $\cO(p^6)$ LECs $C_{i=12,14,17,19,31}$ in Eq.~\eqref{eq.lagnnlo}, we shall take the theoretical estimation from Refs.~\cite{Jiang:2009uf,Jiang:2015dba}. In order to account the uncertainties of the $\cO(p^6)$ LECs, we introduce a common free coefficient $\alpha$ for the $C_i$ terms. The coefficient $\alpha$ will be fitted. 
For the remaining six LECs $L_4, L_6, L_7, L_{18}, L_{25}, v_2^{(2)}$, we shall take two strategies to estimate their values. In one case, $L_4, L_6$ and $L_7$, which exist in the $\cO(p^4)$ $SU(3)$ $\chi$PT Lagrangian, will be fitted. While for $L_{18}, L_{25}, v_2^{(2)}$, that only appear in $U(3)$ $\chi$PT and affect exclusively the quantities involving $\eta$ and $\eta'$ mesons, we fix them at vanishing values, as done in Ref.~\cite{Guo:2015xva}. This strategy is well motivated because the NLO fits are found to be able to qualitatively reproduce the lattice data of the $\eta$-$\eta'$ mixing. In this case we have nine free parameters to fit the 137 data points from several independent lattice simulations and phenomenological inputs. In the other strategy, we try to free all of the six LECs $L_4, L_6, L_7, L_{18}, L_{25}, v_2^{(2)}$ and fit them to the same data sets as the former case. The inclusion of the additional lattice data related to the $\eta$-$\eta'$ mixing from Ref.~\cite{Ottnad:2017bjt}, which are absent in Ref.~\cite{Guo:2015xva}, helps to stabilize the fits with many parameters. 
The conservative bounds for $|\Lambda_1$ and $\Lambda_2$ estimated in Ref.~\cite{Guo:2015xva} and the positive conditions for $L_5$ and $L_8$ from resonance saturations~\cite{Ecker:1988te} also provide useful criteria to obtain meaningful fits.

\begin{table}[htbp]
 \centering
{\footnotesize
\begin{tabular}{c|cc  }
\hline\hline
                      &{NLO Fit  ($F$) }     &{NLO Fit ($F_\pi$) }      \\\hline
$\chi^2/(d.o.f)$      &$471.6/(137-5)$    &$328.9/(137-5)$     \\
$F({\rm MeV})$        &$91.97\pm0.42$      &$91.43\pm0.40$  \\
$10^3\times L_5$      &$1.46 \pm0.04$      &$1.73 \pm0.05$ \\
$10^3\times L_8$      &$0.76 \pm0.04$      &$0.94 \pm0.06$ \\
$\Lambda_1$           &$-0.18\pm0.05$      &$-0.17\pm0.05$ \\
$\Lambda_2$           &$-0.02\pm0.09$      &$0.02 \pm0.09$ \\
\hline\hline
\end{tabular}
\caption{\label{tab.fitnlo} The values of the LECs from the NLO fits. } }
\end{table}

When truncating the $\chi$PT calculation up to a specific finite order, there are always ambiguities to express the physical quantities. E.g., one could use the LO $F$ elsewhere in a physical quantity, or replace $F$ by the renormalized pion decay constant $F_\pi$. In the ideal case when the chiral or $\delta$ expansion works perfectly, one should not expect significant deviations from the two different formalisms, since the differences at least belong to the one order higher effects than the truncated one. In practice, the chiral series may converge slowly, especially for the cases including the strange quark and the $\uao$ anomaly effects. The situations when confronting the lattice simulations with the unphysically large quark masses could become even less clear. E.g. it is mentioned in Ref.~\cite{Bernard:2009ds} that noticeable deviations can appear by expressing the physical quantities with $F$ and $F_\pi$. In Ref.~\cite{Guo:2015xva} the differences between the two schemes are treated as systematical uncertainties, most of which are clearly larger than the statistical ones. In this work we perform different fits by using either $F$ or $F_\pi$ in the expressions of various physical quantities, which are explicitly given in Ref.~\cite{Guo:2015xva}. Regarding the situation of using $F_\pi$ in the final expressions, we employ a slightly different form for the renormalized $F_\pi$, comparing with that in  Ref.~\cite{Guo:2015xva}. In the former reference, the NLO formula of $F_\pi$ was used in the NNLO expressions of the physical quantities, while in this work we use the NNLO formula for $F_\pi$. The difference of the two approaches belongs to a NNNLO effect, which is beyond the accuracy of our current discussion up to NNLO.

\begin{table}[htbp]
 \centering
{\footnotesize
\begin{tabular}{c|cccc  }
\hline\hline
  & {Fit A ($F$) }      &{Fit B ($F$)}    &{Fit A ($F_\pi$) }    &{Fit B ($F_\pi$)}    \\
 \hline
$\chi^2/(d.o.f)$ &$254.0/(137-9)$    &$288.0/(137-9)$   &$292.2/(137-9)$  &$310.7/(137-9)$\\
$F({\rm MeV})$        &$84.15 \pm 3.25$ &$82.79 \pm3.58$ &$92.50 \pm2.74$ &$92.33 \pm2.28$\\
$10^3\times L_5$      &$0.62 \pm0.23$   &$0.48  \pm0.26$ &$1.23  \pm0.26$ &$1.33  \pm0.28$\\
$10^3\times L_8$      &$0.38 \pm0.17$   &$0.39  \pm0.13$ &$0.65  \pm0.16$ &$0.72  \pm0.15$\\
$\Lambda_1$           &$0.17 \pm0.14$   &$0.12  \pm0.13$ &$0.24  \pm0.21$ &$0.29  \pm0.24$\\
$\Lambda_2$           &$0.19 \pm0.19$   &$0.22  \pm0.18$ &$-0.26 \pm0.47$ &$-0.50 \pm0.67$\\
$10^3\times L_4$      &$-0.22 \pm0.13$  &$-0.16 \pm0.14$ &$-0.60 \pm0.13$ &$-0.60 \pm0.10$\\
$10^3\times L_6$      &$-0.20 \pm0.07$  &$-0.08 \pm0.06$ &$-0.33 \pm0.08$ &$-0.25 \pm0.08$\\
$10^3\times L_7$      &$0.33  \pm0.08$  &$0.49  \pm0.09$ &$0.16  \pm0.15$ &$0.20  \pm0.27$\\
$\alpha        $      &$-0.72 \pm0.12$  &$-0.84 \pm0.14$ &$-0.49 \pm0.22$ &$-0.44 \pm0.24$\\
\hline\hline
\end{tabular}
\caption{\label{tab.fit} The values of the LECs from the nine-parameter NNLO fits. The values of the pure $U(3)$ LECs $v_2^{(2)}, L_{18}$ and $L_{25}$ are fixed at zero. The results are quite similar with those in Ref.~\cite{Guo:2015xva}, where the same fit strategies are used. For Fit A, the values of the $\cO(p^6)$ LECs $C_i$ are taken  from Ref.~\cite{Jiang:2009uf}. For Fit B the $C_i$ values are from Ref.~\cite{Jiang:2015dba}. The symbols $F$ or  $F_\pi$ accompanying Fit A and Fit B correspond to the fits using the LO $F$ or the renormalized $F_\pi$ to express the physical quantities. See the text for details. }
}
\end{table}

\begin{table}[htbp]
 \centering
{\footnotesize
\begin{tabular}{c|c|c|c|c }
\hline\hline
  & {Fit A ($F$) 12P }     &{Fit B ($F$) 12P }   &{Fit A ($F_\pi$) 12P}     &{Fit B ($F_\pi$) 12P}    \\ \hline
$\chi^2$              &$246.7/(137-12)$   &$286.8/(137-12)$    &$269.4/(137-12)$  &$276.5/(137-12)$ \\
$F({\rm MeV})$        &$81.35\pm6.84$      &$81.79\pm6.35$   &$92.01\pm4.57$   &$92.47\pm4.80$  \\
$10^3\times L_5$      &$0.45 \pm0.51$      &$0.42 \pm0.49$    &$1.31 \pm0.53$    &$1.46\pm0.60$   \\
$10^3\times L_8$      &$0.27 \pm0.28$      &$0.34 \pm0.21$    &$0.67 \pm0.30$    &$0.75\pm0.32$   \\
$\Lambda_1$           &$0.38 \pm0.75$      &$0.19 \pm0.36$    &$0.19 \pm0.24$    &$0.15\pm0.24$   \\
$\Lambda_2$           &$-0.42\pm1.40$      &$-0.27\pm1.46$    &$-1.68\pm1.18$    &$-1.78\pm1.32$   \\
$10^3\times L_4$      &$-0.11\pm0.25$      &$-0.12\pm0.24$    &$-0.58\pm0.22$    &$-0.61\pm0.23$   \\
$10^3\times L_6$      &$-0.15\pm0.13$      &$-0.05\pm0.10$    &$-0.26\pm0.12$    &$-0.19\pm0.11$   \\
$10^3\times L_7$      &$0.18 \pm0.29$      &$0.30 \pm0.27$    &$-0.17\pm0.39$    &$-0.12\pm0.42$   \\
$\alpha        $      &$-0.76\pm0.29$      &$-0.84\pm0.29$    &$-0.41\pm0.48$    &$-0.27\pm0.55$   \\
$v_2^{(2)}     $      &$0.02 \pm0.05$      &$0.04\pm0.05$     &$0.04 \pm0.04$    &$0.04\pm0.04$   \\
$10^3\times L_{18}$   &$-0.15\pm0.63$      &$-0.05\pm0.41$    &$0.17 \pm0.57$    &$0.34\pm0.60$   \\
$10^3\times L_{25}$   &$-0.16\pm0.51$      &$-0.14 \pm0.50$   &$-0.68\pm0.42$    &$-0.70\pm0.49$   \\
\hline\hline
\end{tabular}
\caption{ The values of the LECs from the twelve-parameter NNLO fits. The $U(3)$ LECs $v_2^{(2)}, L_{18}$ and $L_{25}$ are fitted. For other notations, see Table~\ref{tab.fit} for details. }\label{tab.fita}
}
\end{table}

\begin{table}[htbp]
 \centering
{\footnotesize
\begin{tabular}{c|ccccc  }
\hline\hline
 {Parameters}  & {Inputs}    &{Fit A ($F$)}      &{Fit B($F$)}      &{Fit A($F_\pi$) }     &{Fit B($F_\pi$)}    \\  
 \hline
$F_0(MeV)$          &\quad$118.1 \pm16.5$	&\quad$104.1 \pm 4.3$	&\quad$103.2 \pm 3.5$	&\quad$106.0 \pm 4.4$	&\quad$106.8 \pm 4.0$\\
$F_8(MeV)$          &\quad$133.8 \pm11.1$	&\quad$112.7 \pm 1.3$	&\quad$112.3 \pm 1.2$	&\quad$113.1 \pm 2.1$	&\quad$111.7 \pm 1.9$\\
$\theta_0(Deg)$     &     $-11.0 \pm3.0$	&     $-3.7  \pm 2.4$	&     $-3.3  \pm 4.4$	&     $-7.0  \pm 2.1$  &      $-7.4  \pm 2.1$\\
$\theta_8(Deg)$     &     $-26.7 \pm5.4$	&     $-24.2 \pm 2.1$	&     $-24.4 \pm 4.8$	&     $-26.1 \pm 2.5$	&     $-25.2 \pm 2.7$\\
$m_s/\hat{m}$       &\quad$27.5  \pm3.0$	&\quad$25.1  \pm 1.7$	&\quad$28.3  \pm 0.9$	&\quad$26.6  \pm 1.0$  &\quad$28.9   \pm 0.6$\\
$F_q(MeV)$          &\quad$106.0 \pm11.1^*$ &\quad$87.1  \pm 3.5$	&\quad$85.9  \pm 2.9$	&\quad$89.8  \pm 4.7$  &\quad$91.3   \pm 4.5$\\
$F_s(MeV)$          &\quad$143.8 \pm16.5^*$ &\quad$126.3 \pm 2.0$	&\quad$126.1 \pm 1.9$	&\quad$126.3 \pm 2.9$	&\quad$124.7 \pm 2.5$\\
$\theta_q(Deg)$     &\quad$34.5  \pm5.4^*$	&\quad$41.9  \pm 2.7$	&\quad$41.9  \pm 5.3$	&\quad$39.6  \pm 2.6$  &\quad$40.3   \pm 2.7$\\
$\theta_s(Deg)$     &\quad$36.0  \pm4.2^*$	&\quad$39.3  \pm 2.3$	&\quad$39.5  \pm 4.3$	&\quad$36.7  \pm 2.3$  &\quad$36.7   \pm 2.3$\\
\hline\hline
\end{tabular}
\caption{\label{tab.fit2} Phenomenological results obtained at physical meson masses using the nine-parameter fits in Table~\ref{tab.fit}. The phenomenological inputs of $F_0$, $F_8$, $\theta_0$ and $\eta_8$ are taken from Ref.~\cite{Chen:2014yta}, as done in Ref.~\cite{Guo:2015xva}. The mixing parameters in the quark-flavor basis $F_q$, $F_s$, $\theta_q$ and $\theta_s$ are not phenomenological inputs in the fits, since they can be determined by $F_0$, $F_8$, $\theta_0$ and $\eta_8$. The input ratio of the strange quark mass and the up/down-quark mass $m_s/\hat{m}$ is taken from the FLAG working group in Ref.~\cite{Aoki:2013ldr} and a $10\%$ error bar is introduced as done in Refs.~\cite{Guo:2015xva,Bijnens:2011tb}. } }
\end{table}

In Table~\ref{tab.fitnlo}, we give the values of the LECs from the NLO fits, which turn out to be compatible with those in Ref.~\cite{Guo:2015xva} within uncertainties. The parameters from the NNLO fits are summarized in Tables~\ref{tab.fit} and~\ref{tab.fita}. Since $m_\eta$ and $m_\eta'$ are reasonably described at LO, we fix $M_0=820.0$~MeV determined from the LO fit in the NLO and NNLO discussions. In Table~\ref{tab.fit} we show the results by fixing $L_{18}, L_{25}, v_2^{(2)}$ at vanishing values and in Table~\ref{tab.fita} we give the results by freeing their values in the fits.  For the fits labeled by Fit A, the values of the $\cO(p^6)$ LECs $C_{i=12,14,17,19,31}$ are taken from Ref.~\cite{Jiang:2009uf}. For Fit B, we take the $C_i$ values from Ref.~\cite{Jiang:2015dba}. The symbols $F$ or $F_\pi$ accompanying Fit A and Fit B correspond to using the LO $F$ or the renormalized $F_\pi$ in the expressions of physical quantities.

The first lesson we learn from Table~\ref{tab.fit} is that the fits by taking different $C_{i=12,14,17,19,31}$ values from Refs.~\cite{Jiang:2009uf,Jiang:2015dba} lead to quite comparable results. By contrast, obvious changes of the fitted parameters result from different fits by using $F$ and $F_\pi$ to express the physical quantities. In fact similar problems have been noticed in previous works~\cite{Bernard:2009ds,Guo:2015xva}, specially when the lattice data with unphysically large quark masses are considered. In order to further discriminate the two fit strategies, one possible way is to include more types of data, such as the scattering phase shifts and inelasticities, which is however beyond the scope of present study. In Ref.~\cite{Guo:2015xva} the differences between the two fit strategies using $F$ and $F_\pi$ in the expressions of physical quantities are treated as the systematical uncertainties, which dominate the large error bars of many LECs in that reference. Within uncertainties the values of the $\chi$PT LECs in Table~\ref{tab.fit} are compatible with the previous determinations in Ref.~\cite{Guo:2015xva}, i.e. the numbers in Table 5 of the former reference.  In this work we explicitly give the values resulting from the two fit strategies. As a result, the present uncertainties shown in all the tables and figures correspond to the statistical ones at 1-$\sigma$ level. Due to the reshuffle of the LECs in the $\delta$ counting in $U(3)$ $\chi$PT, it does not allow us to perform a direct comparison of the values of the $SU(3)$ LECs~\cite{Bijnens:2014lea,Aoki:2016frl}. Nevertheless both the phenomenological and lattice determinations prefer small values in magnitudes for $L_4$ and $L_6$ in  Refs.~\cite{Bijnens:2014lea,Aoki:2016frl}, which are in accord with the large $N_C$ expectation. From this point of view, it seems that the results from the Fit $(F)$ strategies in Tables~\ref{tab.fit} and~\ref{tab.fita} are slightly preferred over those from the Fit ($F_\pi$) cases.

Regarding the differences between the results in Tables~\ref{tab.fit} and~\ref{tab.fita},  we do not observe significant improvements by releasing the three LECs $L_{18}, L_{25}, v_2^{(2)}$ in the fits. The fits in Table~\ref{tab.fita} are labeled by Fit ($F$) 12P and Fit ($F_\pi$) 12P, in order to distinguish the nine-parameter fits in Table~\ref{tab.fit} and to highlight the facts that there are twelve free parameters. Though the $\chi^2$ from the twelve-parameter fits are slightly  decreased, the resulting values of $\Lambda_1$ and $\Lambda_2$ are mostly incompatible with the conservative estimates in Ref.~\cite{Guo:2015xva}, especially for the cases of Fit A($F_\pi$) 12P and Fit B($F_\pi$) 12P. It is verified that the resulting plots from the twelve-parameter fits are quite similar with the nine-parameter cases. In order not to overload the figures, we shall show the representative results of Fit A($F$) 12P for the twelve-parameter fits in the following discussions.

\begin{figure}[htbp]
  \centering
\includegraphics[width=0.7\textwidth,angle=-0]{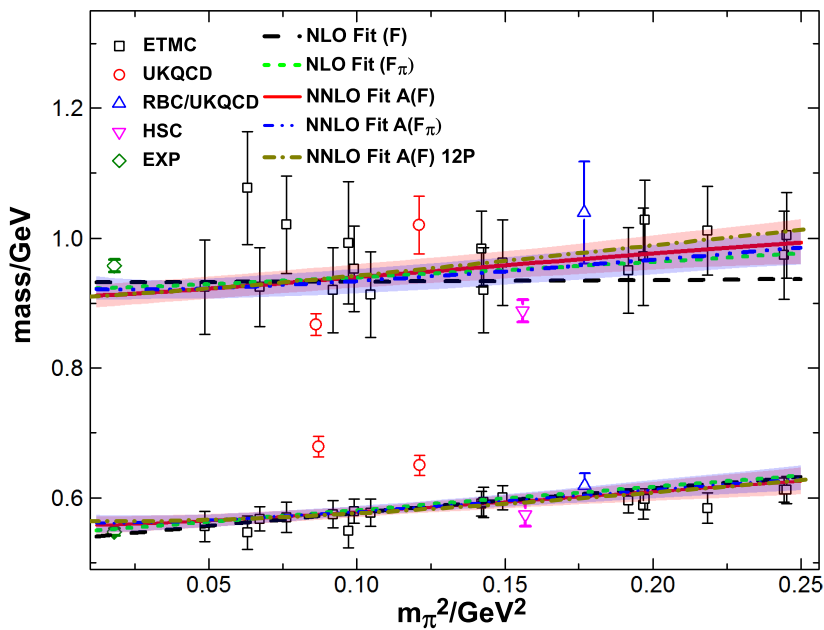}
\caption{ The pion-mass dependences of the masses of the $\eta$ and $\eta'$. The lattice data are taken from Ref.~\cite{Ottnad:2017bjt} (ETMC), Ref.~\cite{ukqcd12prd} (UKQCD), Ref.~\cite{rbcukqcd10prl} (RBC/UKQCD) and Ref.~\cite{hsc11prd} (HSC). For the data from ETMC, including those in Figs.~\ref{fig.fqfs} and \ref{fig.phi}, we have used the results after the corrections of the unphysical strange quark mass and finite lattice spacing. The black long dashed lines and green short dashed lines correspond to the results from NLO Fit ($F$) and NLO Fit ($F_\pi$), respectively. The red solid lines and the surrounding shaded areas denote the central results and 1-$\sigma$ error bars from the NNLO nine-parameter Fit A ($F$). The blue dashed-dotted-dotted lines stand for the results from the NNLO nine-parameter Fit A ($F_\pi$). The results from the NNLO Fit B are similar to those of Fit A and we do not explicitly show them here. As discussed in the text, the twelve-parameter NNLO fits resemble the nine-parameter case. In order not to overload the figure, we only show the representative results from the Fit A ($F$) 12P for the NNLO twelve-parameter fits. Similar rules also apply for the curves in Figs.~\ref{fig.fqfs}-\ref{fig.mk}. }
\label{fig.metametap}
\end{figure}

\begin{figure}[htbp]
   \centering
   \includegraphics[width=0.98\textwidth,angle=-0]{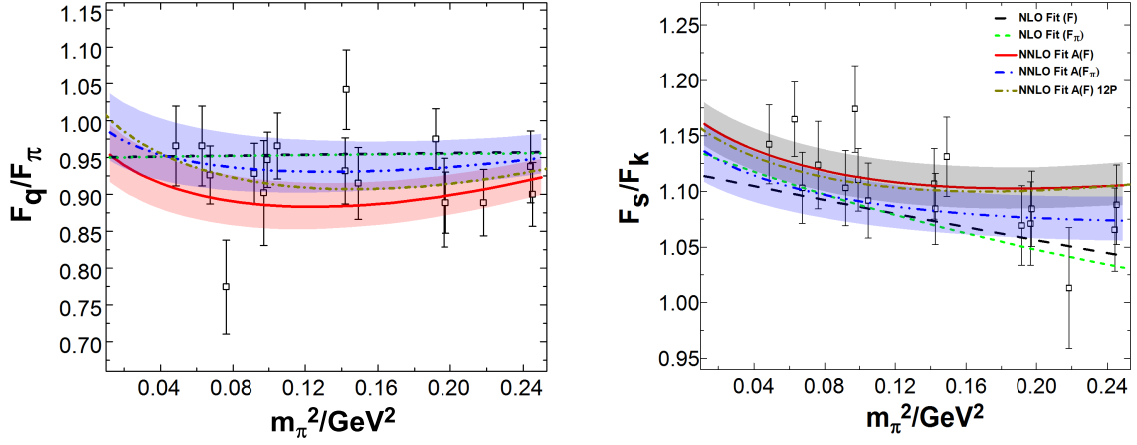}
  \caption{ The pion-mass dependences of the ratios of $F_q/F_\pi$ and $F_s/F_K$, where $F_q$ and $F_s$ are the decay constants  defined in the quark-flavor basis in Eq.~\eqref{twoanglesmixingqs}. The lattice data are taken from Ref.~\cite{Ottnad:2017bjt}. See Fig.~\ref{fig.metametap} for the meaning of the plots.   }
   \label{fig.fqfs}
\end{figure}

\begin{figure}[htbp]
   \centering
   \includegraphics[width=0.7\textwidth,angle=-0]{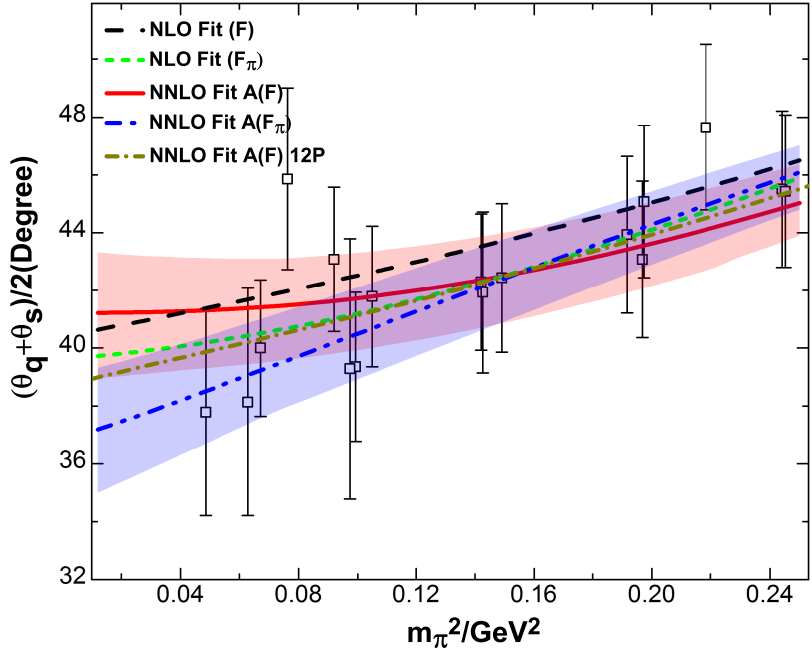}
  \caption{ The pion-mass dependences of the averages of the mixing angles $\theta_q$ and $\theta_s$ are shown, where $\theta_q$ and $\theta_s$ are mixing angles defined in the quark-flavor basis in Eq.~\eqref{twoanglesmixingqs}. The lattice data are taken from Ref.~\cite{Ottnad:2017bjt}.  See Fig.~\ref{fig.metametap} for the meaning of the plots. }
   \label{fig.phi}
\end{figure}

\begin{figure}[htbp]
   \centering
\includegraphics[width=0.98\textwidth,angle=-0]{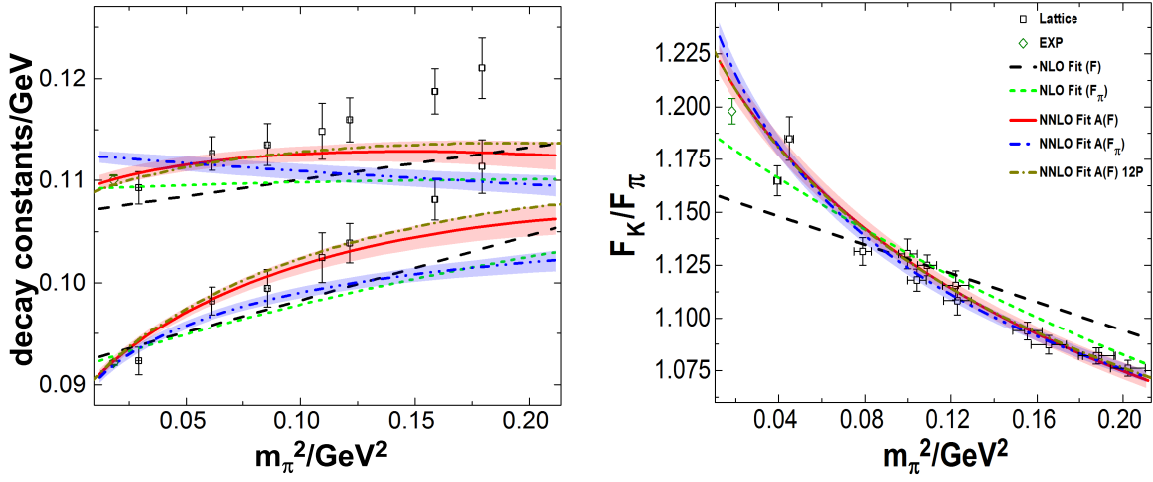}
  \caption{ The pion-mass dependences of the decay constants of the pion and kaon are shown in the left panel, where the lattice data are taken from Refs.~\cite{Aoki:2010dy,Arthur:2012yc}. The pion-mass dependences of the ratio of $F_K/F_\pi$ are shown in the right panel and the corresponding lattice data are taken from Ref.~\cite{Durr:2010hr}. The experimental data (EXP) are taken from Ref.~\cite{Patrignani:2016xqp}. See Fig.~\ref{fig.metametap} for the meaning of the plots. }
   \label{fig.fkfpi}
\end{figure}

\begin{figure}[htbp]
\centering
\includegraphics[width=0.7\textwidth,angle=-0]{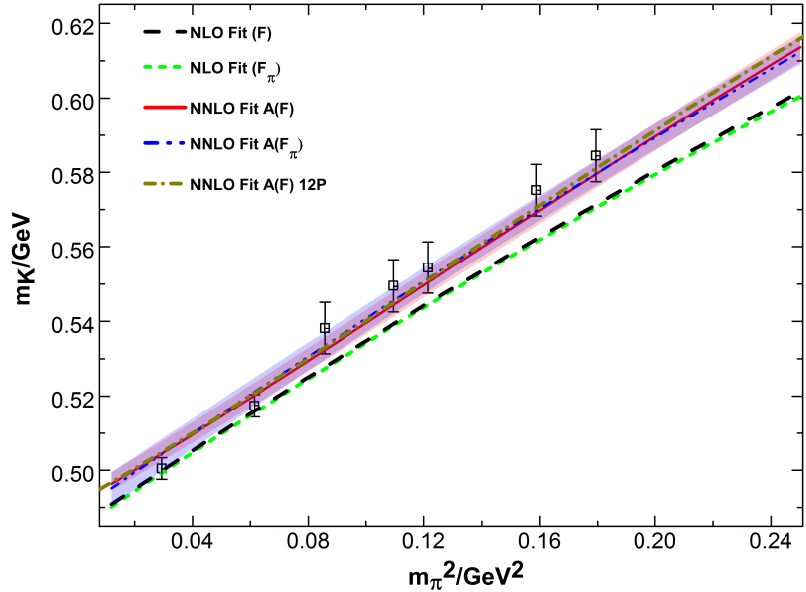}
\caption{ The pion-mass dependences of the kaon masses. The lattice data are taken from Refs.~\cite{Aoki:2010dy,Arthur:2012yc}. See Fig.~\ref{fig.metametap} for the meaning of the plots. }
   \label{fig.mk}
\end{figure}

The phenomenological quantities and the corresponding outputs from the fits are given in Table~\ref{tab.fit2}. In Figs.~\ref{fig.metametap}-\ref{fig.mk}, we show the fit qualities of the lattice simulation data. In Figs.~\ref{fig.metametap}, ~\ref{fig.fqfs} and \ref{fig.phi}, the reproductions of the lattice simulation data of the $\eta$-$\eta'$ mixing are shown. The decay constants of the pion and kaon are given in Fig.~\ref{fig.fkfpi}, where one can see that the NNLO fits considerably improve the fit qualities of the NLO ones. Similar conclusions can be also made for the pion-mass dependences of the kaon masses in Fig.~\ref{fig.mk}. Taking into account the large uncertainties from the lattice simulations on the $\eta$-$\eta'$ mixing, the improvements of the NNLO fits are not obvious, comparing with the NLO ones, see Figs.~\ref{fig.metametap}(masses of $\eta$ and $\eta'$),~\ref{fig.fqfs} and~\ref{fig.phi} (mixing parameters of the $\eta$ and $\eta'$ system). According to these plots in  Fig.~\ref{fig.metametap}-\ref{fig.phi}, we conclude that the formalisms in this work well reproduce the lattice simulation data of the $\eta$ and $\eta'$ mesons. Therefore it gives us a confident starting point to extend our discussions to the finite-temperature case.

\section{Thermal behaviors of the light pseudoscalar mesons}\label{sec.finitetemp}

The masses of the $\pi, K$ and $\eta_8$ mesons are related to the light-flavor quark condensates, which are the order of parameters of spontaneous chiral symmetry breaking. For the massive $\eta_0$, its mass is sensitive to the QCD $\uao$ anomaly effect. The physical $\eta$ and $\eta'$ mesons are the mixture of the $\eta_8$ and $\eta_0$ states. Therefore thermal behaviors of the masses of $\pi, K, \eta$ and $\eta'$ are subject to the restorations of the chiral and $\uao$ symmetries at finite temperatures. For a hadron in the hot medium, its mass is not uniquely defined. The screening mass is usually focused in thermal lattice simulations. In this work we study the thermal behavior of the pole mass, which is extracted from the self-energy of the particle in question and determines its propagation in hot medium.

Within the framework of $\chi$PT, the couplings in front of the local operators are independent of the temperatures and the finite-temperature effects enter through the chiral loops~\cite{Gasser:1986vb,Gerber:1988tt}. In the imaginary time formalism, the extension from $T=0$ to finite $T$ can be achieved by replacing the Minkowski time $t$ with the Euclidean one $\tau=i t$ and then performing the temporal integration along the complex contour $[0, i\beta]$, with $\beta=1/T$. The calculations of the tree-level Feynman diagrams are the same both for $T=0$ and $T\neq 0$. For the loop diagrams, one needs to substitute the continuous integration of the zeroth component $p^0$ of the four momenta by a discrete sum of Matsubara frequencies $i\omega_n=i2\pi n T$, i.e. $\int  d p^0 \to \sum_n i 2\pi T$. As mentioned before, only the tadpole loop diagrams in Fig.~\ref{fig.feynmandiagram}  will enter the present  discussions. Following the standard thermal loop calculation techniques~\cite{Bellacbook}, it is straightforward to obtain the tadpole one-loop function $A_0(m^2)$ at finite $T$,
\begin{eqnarray} \label{eq.a0ft}
 A_0(m^2) = - m^2\ln{\frac{m^2}{\mu^2}} - \int^\infty_0 dp \frac{8p^2}{E_p} \frac{1}{e^{\frac{E_p}{T}}-1}\,,
\end{eqnarray}
with $E_p=\sqrt{p^2+m^2}$. The first term in the right-hand side of Eq.~\eqref{eq.a0ft} corresponds to the tadpole function in vacuum with $T=0$ in Eq.~\eqref{eq.a0} and the finite-temperature effects are introduced via the second term. We mention that at NNLO in the $\delta$ expansion the finite-temperature effects in Eq.~\eqref{eq.a0ft} only contribute to the real part of the self energies of the pNGBs, which will shift the masses from their positions at $T=0$. The imaginary or absorption part starts to appear in the two-loop diagrams, which is beyond the scope of the present study. Notice that the integral in Eq.~\eqref{eq.a0ft} does not have a simple analytical form, but it is straightforward to perform the integration numerically.  To replace the vacuum loop functions in Eq.~\eqref{eq.a0} with the finite temperature corrected ones in Eq.~\eqref{eq.a0ft}, one could study the thermal behaviors of the $\pi, K, \eta$ and $\eta'$ mesons.  Apart from the situation with physical quark masses, we also explore the interesting scenario by varying the quark masses of different flavors at finite temperatures.

\begin{figure}[bhtp]
   \centering
   \includegraphics[width=0.6\textwidth,angle=-0]{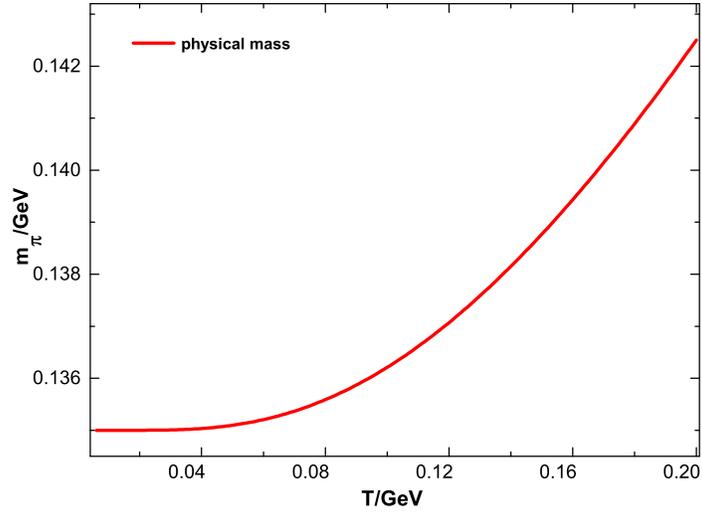}
  \caption{ The temperature dependences of $m_\pi$ with physical quark masses. The curve corresponds to the result from Fit A ($F$). }
   \label{fig.ftmpi}
\end{figure}

\begin{figure}[htbp]
   \centering
   \includegraphics[width=0.98\textwidth,angle=-0]{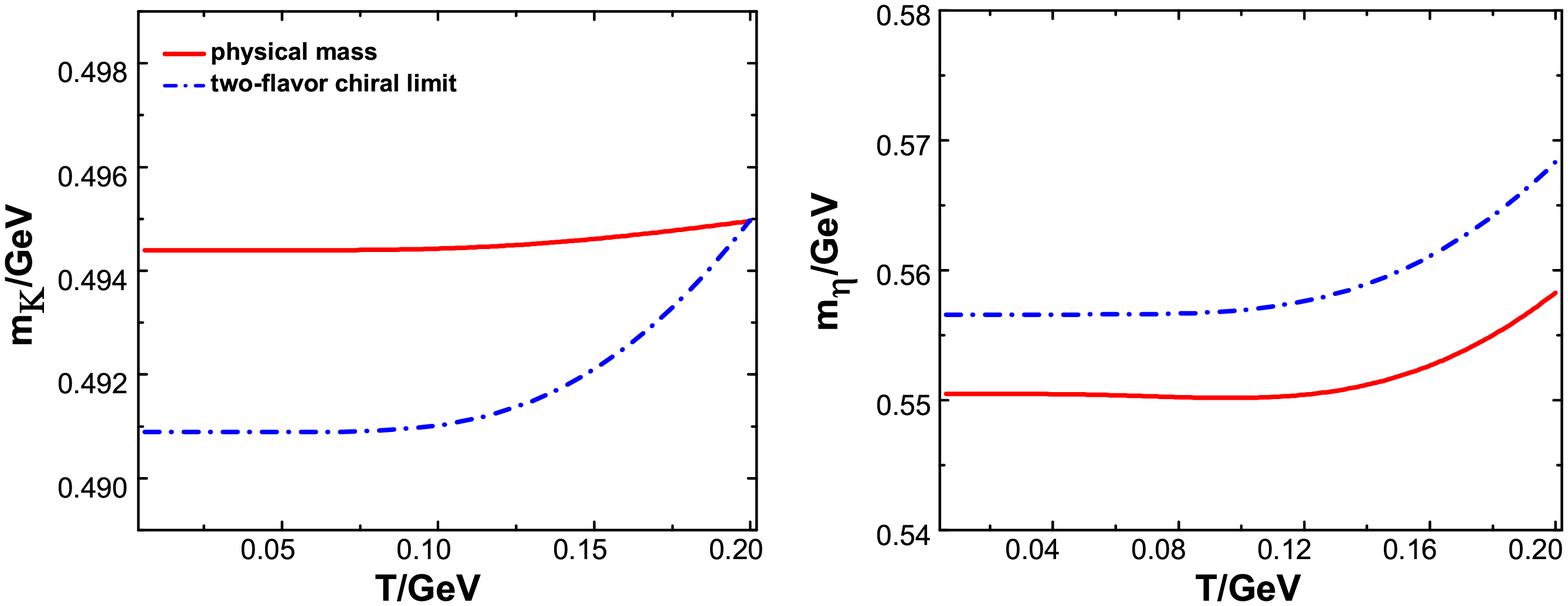}
  \caption{ The temperature dependences of $m_K$ and $m_\eta$. Both the results obtained with physical quark masses and in the two-flavor chiral limit are given. The plots correspond to Fit A ($F$).  }
   \label{fig.ftmkmeta}
\end{figure}

\begin{figure}[htbp]
   \centering
   \includegraphics[width=0.7\textwidth,angle=-0]{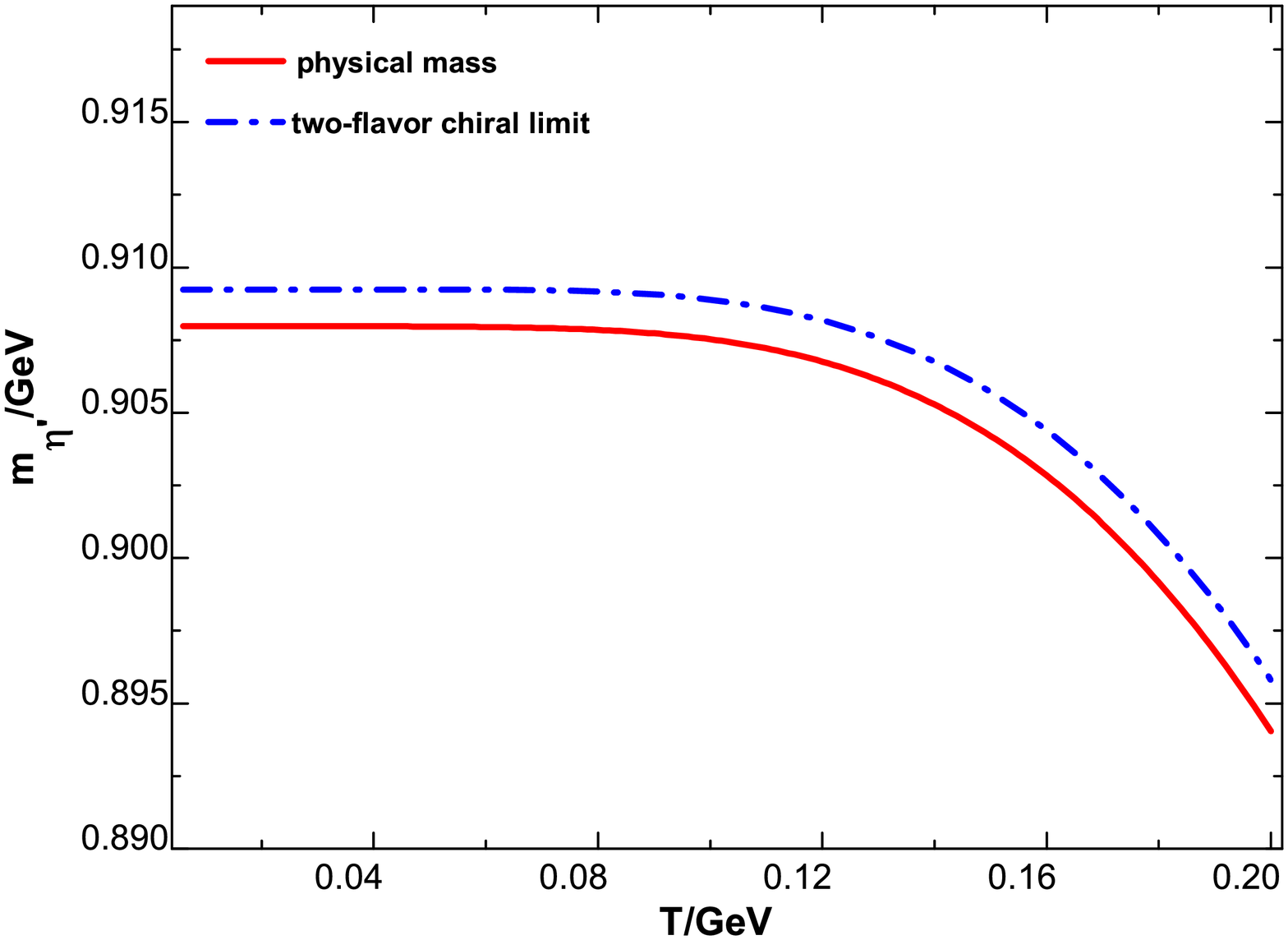}
  \caption{ The temperature dependences of $m_{\eta'}$.  Both the results obtained with physical quark masses and in the two-flavor chiral limit are given. The plots correspond to Fit A ($F$). }
   \label{fig.ftmetap}
\end{figure}

\begin{figure}[htbp]
   \centering
   \includegraphics[width=0.98\textwidth,angle=-0]{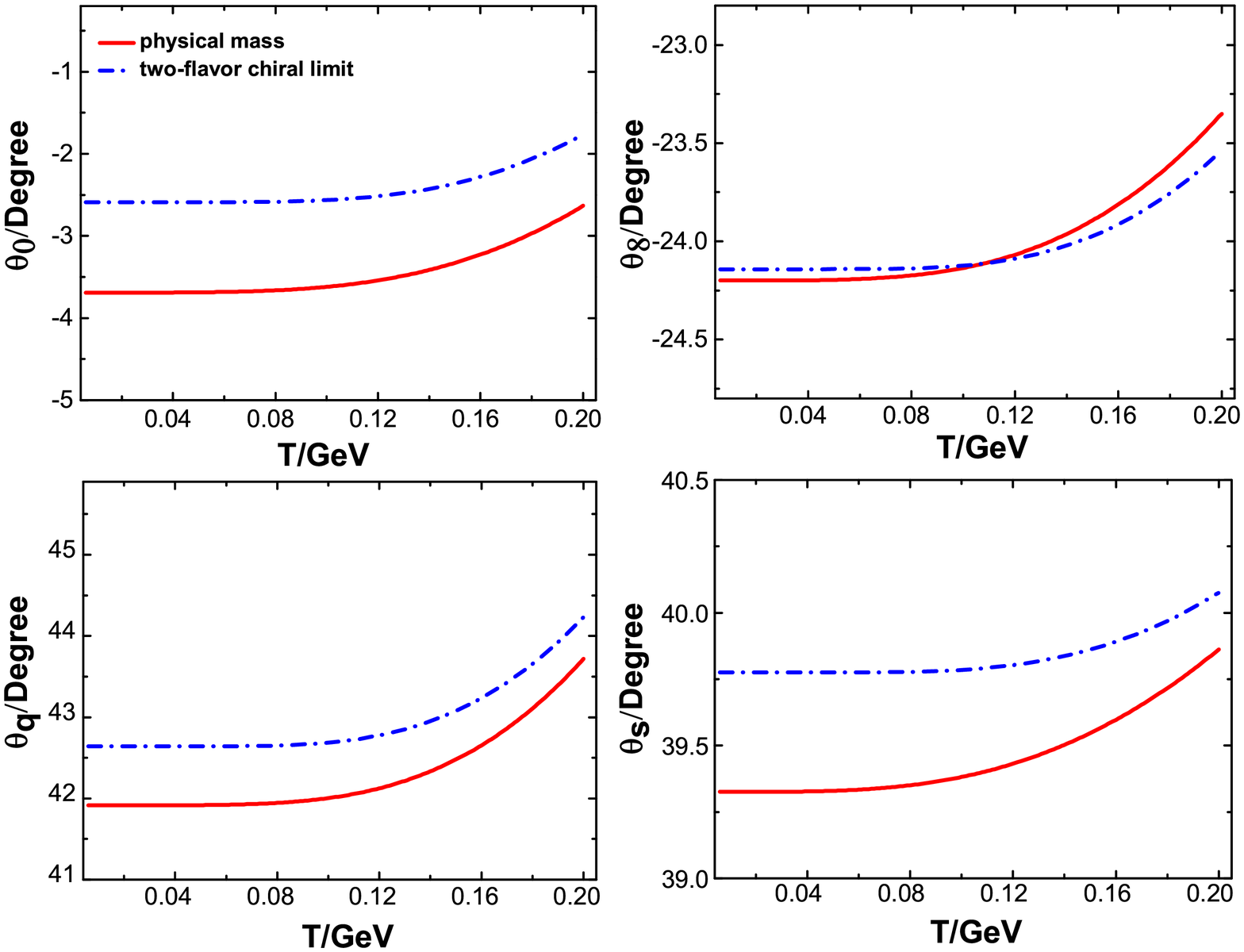}
  \caption{ The temperature dependences of the $\eta$-$\eta'$ mixing angles in the singlet-octet and quark-flavor bases, see Eqs.~\eqref{twoanglesmixing08} and ~\eqref{twoanglesmixingqs}. The results obtained with physical quark masses and in the two-flavor chiral limit are given. The plots correspond to Fit A ($F$).  }
   \label{fig.fttheta}
\end{figure}

The pion decay constant at finite temperature is certainly a model dependent   object~\cite{Gasser:1986vb,Gerber:1988tt,Song:1993ipa,Pisarski:1996mt,Harada:1996pg}.  In order not to muddle the discussions with the model-dependent thermal pion decay constant, we focus on the situation by expressing all the physical quantities in terms of $F$ and refrain from the discussions with the expressions given in terms of the renormalized $F_\pi$. The thermal behaviors of the masses of $\pi,K,\eta$ and $\eta'$ are shown in Figs.~\ref{fig.ftmpi},~\ref{fig.ftmkmeta},~\ref{fig.ftmetap}. The temperature dependences of the mixing angles both in the singlet-octet and quark-flavor bases are given in Fig.~\ref{fig.fttheta}. We emphasize that in $U(3)$ $\chi$PT the thermal corrections to the masses and mixing angles are exclusively contributed by the chiral loops, instead of the chiral LECs. This further implies that the uncertainties of the finite-temperature effects in $\chi$PT are tiny. In order to give concise and intuitive thermal behaviors, we show in Figs.~\ref{fig.ftmpi}-\ref{fig.fttheta} the results  obtained with central values of the LECs from Fit A ($F$) in Table~\ref{tab.fit}. The results from Fit B ($F$) are found to be quite similar with those from Fit A ($F$).

The thermal behavior of the pion masses with physical quark masses is given in Fig.~\ref{fig.ftmpi}. It is obvious that the pion masses are slightly increased when increasing the temperatures. This behavior is consistent with the findings in Refs.~\cite{Gasser:1986vb,Gerber:1988tt}. For the masses of $K, \eta$ and $\eta'$, we study two different scenarios. In one case, we take the physical quark masses and in the other one we take the two-flavor chiral limit, that is to take vanishing $m_{u/d}$ but to keep the  strange quark mass $m_s$ at its physical value. For $m_K$ and $m_\eta$, we find that their masses always get increased when including the finite-temperature contributions in the focused region, no matter the two-flavor chiral limit is taken or not.  In contrast, we observe that the masses of $\eta'$ decrease when the temperatures are increased, both for physical quark masses and the two-flavor chiral limit case. This implies that the meson fluctuation effects in the thermal paths tend to slightly enhance the restoration of the $\uao$ symmetry of QCD, which will also lower the mass of $\eta'$. However the thermal corrections are quite small from the meson fluctuations to the masses. The $\eta$-$\eta'$ mixing angles $\theta_0,\theta_8$ in the singlet-octet basis
and $\theta_q,\theta_s$ in the quark-flavor basis are all increased when increasing the temperatures. The conclusion holds both for physical meson masses and the two-flavor chiral limit case. Similar to the masses, the thermal corrections to the mixing angles turn out to be rather small.

\section{Conclusions}\label{sec.conclusion}

In this work we update the determinations of the low energy constants of the $U(3)$ chiral perturbation theory up to next-to-next-to-leading order, by including phenomenological inputs and various independent lattice simulation data on the $\eta$-$\eta'$ mixing, the kaon masses, the pion and kaon decay constants. Two fit strategies are used in our study. In one case, we closely follow Ref.~\cite{Guo:2015xva} to fix the values of $v_2^{(2)}, L_{18}$ and $L_{25}$ at zero and the fit results are compatible with the previous determinations in the former reference. In the other case, we try to free the values of $v_2^{(2)}, L_{18}$ and $L_{25}$, which turns out to barely improve the fits. The recent lattice simulation data on the $\eta$-$\eta'$ mixing parameters~\cite{Ottnad:2017bjt}, including the masses of $\eta$ and $\eta'$, the  decay constants and the mixing angles in the quark-flavor basis, are well reproduced in our theoretical formalism with reasonable values of the $\chi$PT low energy constants. 

After the successful descriptions of the light pseudo-Nambu-Goldstone bosons in vacuum with $T=0$, we then extend the discussions to finite temperatures with $T \neq 0$. We focus on the thermal behaviors of the masses of $\pi, K, \eta$ and $\eta'$. Up to the next-to-next-to leading order, the finite-temperature effects can only enter through the chiral tadpole loops, which give contributions to the real part of the self-energies of the light pseudo-Nambu-Goldstone bosons. It turns out that at low temperatures the effects from the meson fluctuations slightly increase the masses of $\pi, K,$ and $\eta$ when increasing the temperatures. Interestingly, the mass of the $\eta'$ shows a different behavior from the other mesons and it decreases when increasing the temperatures. This behavior is consistent with the restoration of the $\uao$ symmetry, which will also deduce the mass of the $\eta'$. It indicates that the chiral loops evaluated at finite temperatures slightly enhance the $\uao$ restoration. However the shifts of the thermal masses due to the meson fluctuations turn to be quite small, at most around several percents up to $T=200$~MeV, which cannot account for the mass reduction around $200$~MeV for the $\eta'$ meson~\cite{Csorgo:2009pa}. Another important source is  the finite-temperature effect from the QCD $\uao$ anomaly, which is related to the topological  susceptibility in the gluon sector. Nevertheless the thermal behavior of the pure QCD $\uao$ anomaly could not be accessed in chiral perturbation theory. A future project to also include this effect together with those from the meson fluctuations at finite temperatures, may provide a definite answer to the restoration of the QCD $\uao$ symmetry.

\section*{Acknowledgements}
We would like to thank Jacobo Ruiz de Elvira and Wei-Jie Fu for valuable discussions.
This work is supported in part by the NSFC under Grants
No.~11575052, the Natural Science Foundation of Hebei Province under Contract No.~A2015205205.

\end{document}